\documentstyle[12pt]{article}

\skewchar\fivmi='177
\skewchar\sixmi='177
\skewchar\sevmi='177
\skewchar\egtmi='177
\skewchar\ninmi='177
\skewchar\tenmi='177
\skewchar\elvmi='177
\skewchar\twlmi='177
\skewchar\frtnmi='177
\skewchar\svtnmi='177
\skewchar\twtymi='177
\def\@magscale#1{ scaled \magstep #1}
\font\twfvmi  = ammi10   \@magscale5 
\skewchar\twfvmi='177
\skewchar\fivsy='60
\skewchar\sixsy='60
\skewchar\sevsy='60
\skewchar\egtsy='60
\skewchar\ninsy='60
\skewchar\tensy='60
\skewchar\elvsy='60
\skewchar\twlsy='60
\skewchar\frtnsy='60
\skewchar\svtnsy='60
\skewchar\twtysy='60
\font\twfvsy  = amsy10   \@magscale5 
\skewchar\twfvsy='60

\catcode`@=11
\def\un#1{\relax\ifmmode\@@underline#1\else
        $\@@underline{\hbox{#1}}$\relax\fi}
\catcode`@=12

\let\du=\d                      
\let\um=\H                      

\def\a{\alpha}
\def\b{\beta}

\def\d{\delta}
\def\e{\epsilon}

\def\g{\gamma}

\def\l{\lambda}
\def\m{\mu}
\def\n{\nu}

\def\r{\rho}
\def\s{\sigma}
\def\t{\tau}

\font\sc=font005                        
\def\Sc#1{{\hbox{\sc #1}}}      
\font\ooo=circle10                      
\font\ro=manfnt                         
\def\kcl{{\hbox{\ro 6}}}                
\def\kcr{{\hbox{\ro 7}}}                
\def\ktl{{\hbox{\ro \char'134}}}        
\def\ktr{{\hbox{\ro \char'135}}}        
\def\kbl{{\hbox{\ro \char'136}}}        
\def\kbr{{\hbox{\ro \char'137}}}        

\def\ip{{=\!\!\! \mid}}                                    
\def\bo{{\raise.15ex\hbox{\large$\Box$}}}               
\def\pr{\prod}                                          
\def\TH{{\raise.2ex\hbox{$\displaystyle \bigodot$}\mskip-4.7mu \llap H \;}}
\def\face{{\raise.2ex\hbox{$\displaystyle \bigodot$}\mskip-2.2mu \llap {$\ddot
        \smile$}}}                                      

\def\sp#1{{}^{#1}}                              
\def\Tilde#1{{\widetilde{#1}}\hskip 0.03in}                     
\def\Hat#1{\widehat{#1}}                        
\def\Bar#1{\overline{#1}}                       
\def\leftrightarrowfill{$\mathsurround=0pt \mathord\leftarrow \mkern-6mu
        \cleaders\hbox{$\mkern-2mu \mathord- \mkern-2mu$}\hfill
        \mkern-6mu \mathord\rightarrow$}
\def\dvec#1{\vbox{\ialign{##\crcr
        \leftrightarrowfill\crcr\noalign{\kern-1pt\nointerlineskip}
        $\hfil\displaystyle{#1}\hfil$\crcr}}}           
\def\dt#1{{\buildrel {\hbox{\LARGE .}} \over {#1}}}     

\def\frac#1#2{{\textstyle{#1\over\vphantom2\smash{\raise.20ex
        \hbox{$\scriptstyle{#2}$}}}}}                   
\def\ha{\frac12}                                        
\def\sfrac#1#2{{\vphantom1\smash{\lower.5ex\hbox{\small$#1$}}\over
        \vphantom1\smash{\raise.4ex\hbox{\small$#2$}}}} 
\def\bfrac#1#2{{\vphantom1\smash{\lower.5ex\hbox{$#1$}}\over
        \vphantom1\smash{\raise.3ex\hbox{$#2$}}}}       
\def\afrac#1#2{{\vphantom1\smash{\lower.5ex\hbox{$#1$}}\over#2}}    

\newskip\humongous \humongous=0pt plus 1000pt minus 1000pt
\def\caja{\mathsurround=0pt}
\def\eqalign#1{\,\vcenter{\openup2\jot \caja
        \ialign{\strut \hfil$\displaystyle{##}$&$
        \displaystyle{{}##}$\hfil\crcr#1\crcr}}\,}
\newif\ifdtup
\def\panorama{\global\dtuptrue \openup2\jot \caja
        \everycr{\noalign{\ifdtup \global\dtupfalse
        \vskip-\lineskiplimit \vskip\normallineskiplimit
        \else \penalty\interdisplaylinepenalty \fi}}}
\def\li#1{\panorama \tabskip=\humongous                         
        \halign to\displaywidth{\hfil$\displaystyle{##}$
        \tabskip=0pt&$\displaystyle{{}##}$\hfil
        \tabskip=\humongous&\llap{$##$}\tabskip=0pt
        \crcr#1\crcr}}

\def\ref#1{$\sp{#1)}$}

\parskip=\medskipamount                 
\thicklines                         

\def\oldheadpic{                                
        \setlength{\unitlength}{.4mm}
        \thinlines
        \par
        \begin{picture}(349,16)
        \put(325,16){\line(1,0){4}}
        \put(330,16){\line(1,0){4}}
        \put(340,16){\line(1,0){4}}
        \put(335,0){\line(1,0){4}}
        \put(340,0){\line(1,0){4}}
        \put(345,0){\line(1,0){4}}
        \put(329,0){\line(0,1){16}}
        \put(330,0){\line(0,1){16}}
        \put(339,0){\line(0,1){16}}
        \put(340,0){\line(0,1){16}}
        \put(344,0){\line(0,1){16}}
        \put(345,0){\line(0,1){16}}
        \put(329,16){\oval(8,32)[bl]}
        \put(330,16){\oval(8,32)[br]}
        \put(339,0){\oval(8,32)[tl]}
        \put(345,0){\oval(8,32)[tr]}
        \end{picture}
        \par
        \thicklines
        \vskip.2in}
\def\oldtitle#1#2#3#4{\oldheadpic\begin{center}\vglue.5in{\large\bf #1}\\[.6in]
        {#2}\\[.1in] {\it Department of Physics and Astronomy}\\
        {\it University of Maryland, College Park, MD 20742}\\[.6in]
        Physics Publication \#{#3}\\ {#4}\\[1.5in] {\bf Abstract}\\[.1in]
        \end{center} \begin{quotation}}                 
\def\oldTitle#1#2#3#4#5#6#7{\oldheadpic\begin{center} \vglue .4in
        {\large\bf #1}\\[.4in]
        {#2}\\[.1in] {\it Department of Physics and Astronomy}\\
        {\it University of Maryland, College Park, MD 20742}\\[.1in]
        {#3}\\[.1in] {\it {#4}}\\ {\it {#5}}\\[.4in]
        Physics Publication \#{#6}\\ {#7}\\[.5in] {\bf Abstract}\\[.1in]
        \end{center} \begin{quotation}}                 
\def\border{                                            
        \setlength{\unitlength}{1mm}
        \newcount\xco
        \newcount\yco
        \xco=-24
        \yco=12
        \begin{picture}(140,0)
        \put(\xco,\yco){$\ktl$}
        \advance\yco by-1
        {\loop
        \put(\xco,\yco){$\kcl$}
        \advance\yco by-2
        \ifnum\yco>-240
        \repeat
        \put(\xco,\yco){$\kbl$}}
        \xco=158
        \yco=12
        \put(\xco,\yco){$\ktr$}
        \advance\yco by-1
        {\loop
        \put(\xco,\yco){$\kcr$}
        \advance\yco by-2
        \ifnum\yco>-240
        \repeat
        \put(\xco,\yco){$\kbr$}}
        \put(-20,11){\tiny University of Maryland Elementary Particle
Physics University of Maryland Elementary Particle Physics University of
Maryland Elementary Particle Physics}
        \put(-20,-241.5){\tiny University of Maryland Elementary
Particle Physics University of Maryland Elementary Particle Physics
University of Maryland Elementary Particle Physics}
        \end{picture}
        \par\vskip-8mm}
\def\bordero{                                           
        \setlength{\unitlength}{1mm}
        \newcount\xco
        \newcount\yco
        \xco=-24
        \yco=12
        \begin{picture}(140,0)
        \put(\xco,\yco){$\ktl$}
        \advance\yco by-1
        {\loop
        \put(\xco,\yco){$\kcl$}
        \advance\yco by-2
        \ifnum\yco>-240
        \repeat
        \put(\xco,\yco){$\kbl$}}
        \xco=158
        \yco=12
        \put(\xco,\yco){$\ktr$}
        \advance\yco by-1
        {\loop
        \put(\xco,\yco){$\kcr$}
        \advance\yco by-2
        \ifnum\yco>-240
        \repeat
        \put(\xco,\yco){$\kbr$}}
        \put(-20,12){\ooo
bacdefghidfghghdhededbihdgdfdfhhdheidhdhebaaahjhhdahba
hgdedge
   hgfdiehhgdigicba}
        \put(-20,-241.5){\ooo
ababaighefdbfghgeahgdfgafagihdidihiidhiagfedhadbfd
ecdcdfa
   gdcbhaddhbgfchbgfdacfediacbabab}
        \end{picture}
        \par\vskip-8mm}
\def\headpic{                                           
        \indent
        \setlength{\unitlength}{.4mm}
        \thinlines
        \par
        \begin{picture}(29,16)
        \put(165,16){\line(1,0){4}}
        \put(170,16){\line(1,0){4}}
        \put(180,16){\line(1,0){4}}
        \put(175,0){\line(1,0){4}}
        \put(180,0){\line(1,0){4}}
        \put(185,0){\line(1,0){4}}
        \put(169,0){\line(0,1){16}}
        \put(170,0){\line(0,1){16}}
        \put(179,0){\line(0,1){16}}
        \put(180,0){\line(0,1){16}}
        \put(184,0){\line(0,1){16}}
        \put(185,0){\line(0,1){16}}
        \put(169,16){\oval(8,32)[bl]}
        \put(170,16){\oval(8,32)[br]}
        \put(179,0){\oval(8,32)[tl]}
        \put(185,0){\oval(8,32)[tr]}
        \end{picture}
        \par\vskip-6.5mm
        \thicklines}
\def\title#1#2#3#4{\border\headpic {\hbox to\hsize{#4 \hfill UMDEPP #3}}\par
        \begin{center} \vglue .5in {\large\bf #1}\\[.6in]
        {#2}\\[.1in] {\it Department of Physics and Astronomy}\\
        {\it University of Maryland, College Park, MD 20742}\\[1.5in]
        {\bf Abstract}\\[.1in] \end{center} \begin{quotation}}  
\def\Title#1#2#3#4#5#6#7{\border\headpic
        {\hbox to\hsize{#7 \hfill UMDEPP #6}}\par
        \begin{center} \vglue .4in {\large\bf #1}\\[.4in]
        {#2}\\[.1in] {\it Department of Physics and Astronomy}\\
        {\it University of Maryland, College Park, MD 20742}\\[.1in]
        {#3}\\[.1in] {\it {#4}}\\ {\it {#5}}\\[.5in] {\bf Abstract}\\[.1in]
        \end{center} \begin{quotation}}                 
\def\endtitle{\end{quotation}\newpage}                  

\def\sect#1{\bigskip\medskip \goodbreak \noindent{\bf {#1}} \nobreak \medskip}
\def\refs{\sect{References} \footnotesize \frenchspacing \parskip=0pt}
\def\Item{\par\hang\textindent}

\begin{document}
\def\sqrtrf{{\sqrt{1-\fracmm1{r^4}}}}
\def\sinhth{\sinh\vartheta}
\def\coshth{\cosh\vartheta}
\def\sinhthsq{\sinh^2\vartheta}
\def\coshthsq{\cosh^2\vartheta}
\def\onemi{1-{\fracmm1{r^4}}}

\def\gg{{\hbox{\sc g}}}
\def\nt{$~N=2$~}
\def\gg{{\hbox{\sc g}}}
\def\nt{$~N=2$~}
\def\tr{{\rm tr}}
\def\Tr{{\rm Tr}}
\def\mpl#1#2#3{Mod.~Phys.~Lett.~{\bf A{#1}} (19{#2}) #3}

\def\scst{\scriptstyle}
\def\itrema{$\ddot{\scriptstyle 1}$}
\def\Bo{\bo{\hskip 0.03in}}
\def\lrad#1{ \left( A {\buildrel\leftrightarrow\over D}_{#1} B\right) }

\def\ula{{\underline a}} \def\ulb{{\underline b}} \def\ulc{{\underline c}}
\def\uld{{\underline d}} \def\ule{{\underline e}} \def\ulf{{\underline f}}
\def\ulg{{\underline g}} \def\ulm{{\underline m}}
\def\uls{{\underline s}}
\def\uln{{\underline n}}
\def\ulp{{\underline p}} \def\ulq{{\underline q}} \def\ulr{{\underline r}}

\def\plpl{{+\!\!\!\!\!{\hskip 0.009in}{\raise -1.0pt\hbox{$_+$}}
{\hskip 0.0008in}}}

\def\mimi{{-\!\!\!\!\!{\hskip 0.009in}{\raise -1.0pt\hbox{$_-$}}
{\hskip 0.0008in}}}

\def\items#1{\\ \item{[#1]}}
\def\ul{\underline}
\def\un{\underline}
\def\-{{\hskip 1.5pt}\hbox{-}}

\def\kd#1#2{\d\du{#1}{#2}}
\def\fracmm#1#2{{{#1}\over{#2}}}
\def\footnotew#1{\footnote{\hsize=6.5in {#1}}}

\def\low#1{{\raise -3pt\hbox{${\hskip 1.0pt}\!_{#1}$}}}

\def\ip{{=\!\!\! \mid}}
\def\unb{{\underline {\bar n}}}
\def\upb{{\underline {\bar p}}}
\def\um{{\underline m}}
\def\up{{\underline p}}
\def\Phib{{\Bar \Phi}}
\def\Phit{{\tilde \Phi}}
\def\Phibt{{\tilde {\Bar \Phi}}}
\def\Db{{\Bar D}_{+}}
\def\gg{{\hbox{\sc g}}}
\def\nt{$~N=2$~}

\border\headpic {\hbox to\hsize{September 1992 \hfill UMDEPP 93--52}}\par
\begin{center}
\vglue .25in

{\large\bf Exact Solutions for} \\
{\large\bf Self--Dual Yang--Mills and Self--Dual Tensor Multiplets}
{\large\bf on Gravitational Instanton Background}
$\,$\footnote{This
work is supported in part by NSF grant \# PHY-91-19746.} \\[.1in]

\baselineskip 10pt

\vskip 0.25in

Hitoshi NISHINO \\[.2in]
{\it Department of Physics} \\ [.015in]
{\it University of Maryland at College Park}\\ [.015in]
{\it College Park, MD 20742-4111, USA} \\[.1in]
and\\[.1in]
{\it Department of Physics and Astronomy} \\[.015in]
{\it Howard University} \\[.015in]
{\it Washington, D.C. 20059, USA} \\[.18in]

\vskip 1.0in

{\bf Abstract}\\[.1in]

\end{center}

\begin{quotation}

        We give exact solutions for a recently developed ~$N=1$~ locally
supersymmetric self-dual gauge theories in $~(2+2)\-$dimensions.  We give the
exact solutions for an $~SL(2)$~ self-dual Yang-Mills multiplet and
what we call ``self-dual tensor multiplet'' on the
gravitational instanton background by Eguchi-Hanson.
We use a general method to get an $~SL(2)$~ self-dual Yang-Mills
solution from any known self-dual gravity solution.
Our result is the first example of exact
solutions for the coupled system of these $~N=1$~ locally supersymmetric
self-dual multiplets in ~$(2+2)\-$dimensions,
which is supposed to have strong significance for integrable models
in lower-dimensions upon appropriate dimensional reduction or
truncation.  We also inspect the consistency of our exact solutions as
a background for $~N=2$~ superstring coupled to the Wess-Zumino-Witten
term in $~\s\-$model formulation.

\endtitle

\def\doit#1#2{\ifcase#1\or#2\fi}
\def\[{\lfloor{\hskip 0.35pt}\!\!\!\lceil}
\def\]{\rfloor{\hskip 0.35pt}\!\!\!\rceil}
\def\delsl{{{\partial\!\!\! /}}}
\def\caldsl{{\calD\!\!\! /}}
\def\calO{{\cal O}}
\def\asym{({\scriptstyle 1\leftrightarrow \scriptstyle 2})}
\def\Lag{{\cal L}}
\def\du#1#2{_{#1}{}^{#2}}
\def\ud#1#2{^{#1}{}_{#2}}
\def\dud#1#2#3{_{#1}{}^{#2}{}_{#3}}
\def\udu#1#2#3{^{#1}{}_{#2}{}^{#3}}
\def\calD{{\cal D}}
\def\calM{{\cal M}}
\def\tildef{{\tilde f}}
\def\calDsl{{\calD\!\!\!\! /}}

\def\Hat#1{{#1}{\large\raise-0.02pt\hbox{$\!\hskip0.038in\!\!\!\hat{~}$}}}
\def\hati{{\hat{I}}}
\def\dt{$~D=10$~}
\def\alp{\alpha{\hskip 0.007in}'}
\def\oalp#1{\alp^{\hskip 0.007in {#1}}}
\def\naive{{{na${\scriptstyle 1}\!{\dot{}}\!{\dot{}}\,\,$ve}}}
\def\items#1{\vskip 0.05in\Item{[{#1}]}}
\def\item#1{\Item{#1}}

\def\pl#1#2#3{Phys.~Lett.~{\bf {#1}B} (19{#2}) #3}
\def\np#1#2#3{Nucl.~Phys.~{\bf B{#1}} (19{#2}) #3}
\def\prl#1#2#3{Phys.~Rev.~Lett.~{\bf #1} (19{#2}) #3}
\def\pr#1#2#3{Phys.~Rev.~{\bf D{#1}} (19{#2}) #3}
\def\cqg#1#2#3{Class.~and Quant.~Gr.~{\bf {#1}} (19{#2}) #3}
\def\cmp#1#2#3{Comm.~Math.~Phys.~{\bf {#1}} (19{#2}) #3}
\def\jmp#1#2#3{Jour.~Math.~Phys.~{\bf {#1}} (19{#2}) #3}
\def\ap#1#2#3{Ann.~of Phys.~{\bf {#1}} (19{#2}) #3}
\def\prep#1#2#3{Phys.~Rep.~{\bf {#1}C} (19{#2}) #3}
\def\ptp#1#2#3{Prog.~Theor.~Phys.~{\bf {#1}} (19{#2}) #3}
\def\ijmp#1#2#3{Int.~Jour.~Mod.~Phys.~{\bf {#1}} (19{#2}) #3}
\def\nc#1#2#3{Nuovo Cim.~{\bf {#1}} (19{#2}) #3}
\def\ibid#1#2#3{{\it ibid.}~{\bf {#1}} (19{#2}) #3}

\def\szet{{${\scriptstyle \b}$}}
\def\ula{{\un a}}
\def\ulb{{\un b}}
\def\ulc{{\un c}}
\def\uld{{\un d}}
\def\ulA{{\un A}}
\def\ulM{{\underline M}}
\def\cdm{{\Sc D}_{--}}
\def\cdp{{\Sc D}_{++}}
\def\vTheta{\check\Theta}
\def\Pisl{{\Pi\!\!\!\! /}}

\def\fracmm#1#2{{{#1}\over{#2}}}
\def\gg{{\hbox{\sc g}}}
\def\half{{\fracm12}}
\def\ha{\half}

\def\frac#1#2{{\textstyle{#1\over\vphantom2\smash{\raise -.20ex
        \hbox{$\scriptstyle{#2}$}}}}}                   

\def\fracm#1#2{\hbox{\large{${\frac{{#1}}{{#2}}}$}}}

\def\Dot#1{\buildrel{_{_{\hskip 0.01in}\bullet}}\over{#1}}
\def\dt#1{\Dot{#1}}
\def\uln{{\underline n}}
\def\Tilde#1{{\widetilde{#1}}\hskip 0.015in}
\def\Hat#1{\widehat{#1}}
\def\Dot#1{\buildrel{_{_{\hskip 0.01in}\bullet}}\over{#1}}
\def\dt#1{\Dot{#1}}
\def\uln{{\underline n}}
\def\Tilde#1{{\widetilde{#1}}\hskip 0.015in}
\def\Hat#1{\widehat{#1}}

\oddsidemargin=0.03in
\evensidemargin=0.01in
\hsize=6.5in
\textwidth=6.5in

\noindent 1.{\it ~~Introduction.}~~Recently there has been important
observation in $~N=2$~ superstring theory [1] that the massless background
fields for $~N=2$~ superstring are to be the self-dual Yang-Mill (SDYM)
[2] or self-dual gravity (SDG) fields.  Since the space-time
supersymmetry is to be built-in in such superstring theory, it is natural
to expect that these background fields have to be also {\it supersymmetric}.
Motivated by this development, we have constructed in our recent papers [3-6]
self-dual {\it supersymmetric} YM (SDSYM) theories and
self-dual supergravity (SDSG) theories in four-dimensional
space-time with the signature $~(+,+,-,-)$.\footnotew{We denote this
space-time by $~D=(2,2)$.}~~Another strong
motivation to study SDYM theory is from the conjecture [7]
that {\it all} exactly soluble (bosonic) models in
lower-dimensions can be obtained from the SDYM theory [8].  Moreover the
$~W_{\infty}\-$algebra [9] is also likely to be connected to the
SDG theory in $~D=(2,2)$.

        According to a more recent analysis [10], open $~N=2$~
superstring allows only $~N=4~$ SDSYM, while closed
$~N=2$~ superstring allows only $~N=8$~ SDSG as consistent
target space-time backgrounds, provided that the
background is described by a single irreducible superfield.  Even if
these ``maximal'' supersymmetries may be singled out by consistency for
the irreducibility in the $~N=2$~
superstring, it will be still important to consider some truncated
supersymmetries of these backgrounds, such as $~N\le 4$~ SDSG [5] or $~N\le
2$~ SDSYM [3,5], from the viewpoint of soluble systems in
lower-dimensions [7].

        In our recent paper [6] we gave an exact solution for the gaugino
field in the {\it global} \hbox{$~N=2$}~ SDSYM theory on the bosonic
YM instanton background.  We have seen that the gaugino solution is generated
by a {\it globally} supersymmetric transformation of the gaugino on the bosonic
instanton background.

        In this Letter, we give exact solutions for the $~N=1$~ SDYM
multiplet for the gauge group $~SL(2) \approx SO(1,2)$~
coupled to what we call ``self-dual tensor multiplet''
(SDTM) and the $~N=1$~ SDSG on the space-time background of Eguchi-Hanson
instanton metric [11].  Since our method is based on the general feature of the
SDG and SDYM system, it will also provide a general algorithm generating exact
solutions for $~SL(2)$~ SDYM, whenever a purely gravitational solution
for the SDG is given.

\bigskip\bigskip

\noindent 2.{\it ~~Field Equations.}~~We first review our relevant field
equations in the system.  The field content of the $~D=(2,2),\,N=1$~ SDSG is
$~(e\du\m m,\, \Tilde\psi\du \m{\Dot\a})$, that of the SDSYM is $~(A\du \m I,\,
\l\du\a I)$~ and that of the SDTM is $~(B_{\m\n}, \,\Phi, \,\chi_\a)$.
The $~\Phi$~ is the usual dilaton, while $~B_{\m\n}$~ couples to
the $~N=2$~ superstring [4] {\it via} the Wess-Zumino-Witten term.
In this paper we use the indices $~{\scst \m,~\n,~\cdots~=~1,~\cdots,~4}$~
for the {\it curved} world coordinates, and $~{\scst
m,~n,~\cdots~=~1,~\cdots,~4}$~ for the local Lorentz coordinates.  For
the spinors we use the same convention as in Refs.~[3-6], namely $~{\scst
\a,~\b,~\cdots~=~1,~2}$~ and $~{\scst \Dot\a,~\Dot\b,~\cdots~=~\Dot1,~
\Dot2}$~ are respectively for the {\it chiral} and {\it anti-chiral}
components.  All the {\it anti-chiral} spinors are denoted by the {\it
tilde}, such as $~\Tilde\psi\du\m{\Dot\a}$.  We use the $~2\times2$~
matrices $~(\g^\m)_{\a\Dot\b}$~ and $~(\g^\m)_{\Dot\a\b}$~ instead of
$~(\s^\ulm)_{\a\Dot\b}$~ and $~(\Tilde\s^\ulm)_{\Dot\a\b}$~ in Refs.~[3-6].
The indices $~{\scst I,~J,~\cdots~=~ 1,~2,~3}$~ are for the adjoint
representations for the $~SL(2)~$ gauge group.

        To get the exact solutions, we use the {\it canonical}
set\footnotew{The word {\it canonical} comes from the fact that the {\it
kinetic} terms in the original {\it non-self-dual} lagrangian before
imposing the SD conditions have the standard coefficients.  The
difference between {\it canonical} and non-canonical versions is just a
matter of field-redefinitions, but in practice the former set has such an
advantage as the vanishing torsion, {\it etc.}  For other details, see
Ref.~[12].} of field equations [12], because it has {\it no}
torsion making the self-duality (SD) of the Riemann tensor manifest.
One important point about the supersymmetric self-dual system in general
is that the bosonic field
equations in our system stay exactly the same as the {\it non-supersymmetric}
theory, especially for the SDSYM and SDSG fields, while the bosonic fields
for the SDTM satisfy the peculiar field equations:
$$\li{& i (\g^\n)_{\a\Dot\b}\Tilde T\du{\m\n} {\Dot\b} = 0 ~~,
&(2.1) \cr
&i (\g^\m)\ud\b{\Dot\a} D_\m\chi{\low\b}
+ \fracm i {2\sqrt3} (\g^\m\chi)
_{\Dot\a} \partial_\m\Phi - \fracm1{2\sqrt3} (\g^{\m\n}\Tilde\l^I)_{\Dot\a}
e^{- \Phi/\sqrt3} F\du{\m\n} I = 0~~,
&(2.2) \cr
& i(\g^\m)\du\a{\Dot\b} D_\m\Tilde\l _{\Dot\b}{}^I
- \fracm i{2\sqrt3} (\g^\m \Tilde\l^I)_\a \partial_\m \Phi = 0 ~~,
&(2.3) \cr
& R^{\m\n\r\s} = \half e^{-1} \e^{\m\n\t\omega}
R\du{\t\omega}{\r\s} ~~,~~~~R_{\m\n} = 0~~,
&(2.4) \cr
& \Bo \Phi + \fracm2{\sqrt3} (\partial_\m\Phi)^2 - \fracm1{2\sqrt3}
e^{-2\Phi/\sqrt3} F_{\m\n}{}^I F\ud{\m\n} I = 0~~, ~~~~
G^{\m\n\r} = e^{2\Phi/{\sqrt3}}\, e^{-1} \e^{\m\n\r\s}
\,\partial_\s \Phi ~~,{\hskip 0.08in}
&(2.5) \cr
& F^{\m\n\, I} = \half e^{-1} \e^{\m\n\r\s} F\du{\r\s} I
{}~~, ~~~~ D_\m (e F^{\m\n\, I}) = 0 ~~.
&(2.6) \cr} $$
The derivative $~D_\m$~ is both Lorentz and gauge-covariant, and
$~\Tilde T\du{\m\n} {\Dot\b}$~ is the gravitino field strength.
The field strength $~G_{\m\n\r}$~ of $~B_{\m\n}$~ contains the
Chern-Simons (CS) term:
$$G_{\m\n\r} \equiv 3\partial_{\[ \m} B_{\n\r\]} - {\sqrt3} \left[
F\du {\[\m\n} I A_{\r\]\, I} - \fracm 13 f\du{I J} K A\du\m I A\du \n J
A_{\r K} \right]~~.
\eqno(2.7) $$
Here the antisymmetrizations with the symbols $~{\scst \[~\]}$~
are always {\it normalized}.
The special factor with $~{\sqrt3}$~ comes from the normalization in our
{\it canonical} set of field equations [12].  The appearance of the special
factors like $~e^{2\Phi/\sqrt3}$~ is also due to our canonical system.
The second equation in (2.5) implies
a ``generalized'' SD, namely the third-rank
tensor $~G_{\m\n\r}$~ is {\it dual} to the first-rank field strength
$~\partial_\m \Phi$.
In the above field equations all the
field strengths need {\it no supercovariantization} due to the
special property of the self-dual supersymmetry.  For example, the
field strength $~F\du{\m\n} I$~ has {\it no}
gravitino-dependent term, because the gravitino $~\Tilde\psi\du\m
{\Dot\a}$~ lacks its {\it chiral} partner $~\psi\du\m\a$, so that there
is no possibility such as $~\psi\g \Tilde\l\-$terms.
Since $~SL(2)~$ is a non-compact group, we always need its metric
$~g{\low{I J}} = \hbox{diag.}\,(1,-1,-1)$~ for the contractions of
the indices $~{\scst I,~J,~\cdots~=~1,~2,~3}$.

\bigskip\bigskip

\noindent 3.{\it ~~Exact Solutions.}~~We now give the exact solutions
for our field equations (2.1) - (2.6).  We start with the field
equations (2.1) and (2.4) for the SDSG.  Eq.$\,$(2.4) is satisfied by what
is called Eguchi-Hanson (EH) gravitational instanton solution [11],
modified for our $~D=(2,2)$.  Such EH metric is given by
$$\eqalign{&g_{1 1} = \fracmm1{\onemi} ~~, ~~~~ g_{2 2} = \fracmm{r^2} 4
\left( 1 - \fracmm1{r^4} \coshthsq \right)~~, ~~~~
g_{2 4} = \fracmm 1{r^4} \left( \onemi \right)\coshth~~, \cr
&g_{3 3} = - \fracmm14 r^2 ~~, ~~~~ g_{44} = \fracmm{r^2}4
\left( \onemi \right) ~~, \cr }
\eqno(3.1) $$
where our choice of coordinates is $~(x^\m) =
(r,\,\varphi,\,\vartheta,\,\psi)$, and $~0 \le \varphi < 2\pi,~
0 \le \vartheta <\infty,~ 0 \le \psi < 2\pi$.

        The gravitino equation (2.1) is satisfied by the trivial solution
$~\Tilde\psi\du\m {\Dot\a} = 0$, and this does {\it not} pose any
problem for the following reason.
First, we can easily show that the {\it pure-gauge} solution
$$\Tilde\psi\du\m{\Dot\a} = D_\m \Tilde\l^{\Dot\a} ~~,
\eqno(3.2) $$
with an arbitrary space-time dependent spinor $~\Tilde\l$~ together with
the EH background (3.1) satisfies our field equations (2.1), due to
the identity $~R_{\[\m\n\r\]}{}^\s \equiv 0$.  Now recall that
the supertranslation rule
$$\d\Tilde\psi\du\m{\Dot\a} = D_\m \Tilde\e^{\Dot\a} ~~, ~~~~
\d e\du\m m = - i (\e \g^m \Tilde\psi_\m) ~~,
\eqno(3.3) $$
which can completely gauge away the above solution (3.2).  This means
that by choosing an appropriate frame of supersymmetry, we can put the
background of the gravitino $~\Tilde\psi_\m $~ to be zero.\footnotew{Of
course, however, this does {\it not} exclude the existence of other
gauge-non-trivial solutions for the gravitino.  Our choice is just
one choice of gauge-trivial family of exact solutions for the gravitino.  Other
non-trivial gravitino solutions are yet to be studied in the
future.}~~Therefore we simply set
the gravitino to be zero from now on.

        Our next task is to solve the SDSYM field equations (2.3) and
(2.6).  We can easily find a non-trivial solution to (2.6) on our
EH background (3.1).  The strategy is to utilize the fact that we can
choose a gauge group $~SL(2)$, which coincides with one of the subgroups
of the Lorentz group in the SDG: $~SO(2,2) \approx SL(2)
\otimes SL(2)$.  This method has been known for the Euclidean case for
getting a SDYM solution for the gauge group $~SU(2)$~ out of any known SDG
solution in the Euclidean space-time [13].

        To be more specific, we can identify the Lorentz connection
$~\omega\du\m{m n}$~ for the EH instanton background (3.1) with our YM
gauge field $~A\du\m I$~ as
$$ \eqalign{&\omega\du\m{(1)(2)} = \omega\du\m {(3)(4)}
\rightarrow A_\m{}^1 ~~, \cr
&\omega\du\m{(1)(3)} = \omega\du\m{(2)(4)} \rightarrow A\du\m 2 ~~, \cr
& \omega\du\m{(1)(4)} = \omega\du\m{(3)(2)}\rightarrow A\du\m 3 ~~. \cr }
\eqno(3.4) $$
Here we use the indices in the parentheses such as $~{\scst (1),
{}~\cdots,~(4)}$~
for the {\it flat} Lorentz indices $~{\scst m,~n,~\cdots}$,
distinguished from the {\it curved} ones $~{\scst
\m,~\n,~\cdots~=~1,~\cdots,~4}$~ without parentheses.
Notice that this identification has been made possible, owing to
the manifest SD for the $~{\scst m n}\-$indices of
$~\omega\du\m{m n}$.\footnotew{Tish is {\it not} generally true for any
Lorentz connection for a given self-dual Riemann tensor [14].}~~The
$~SL(2)$~ gauge roup has the generators $~T_I$~ satisfying
$$ \[ T_1,\, T_2\] = - 2T_3~~, ~~~~ \[T_2,\,T_3\] = + 2 T_1~~, ~~~~
\[ T_3,\,T_1\] = -2T_2 ~~.
\eqno(3.5) $$
Relevantly, we can rewrite them in terms of the familiar Virasoro algebra
notation:
$$ \[ L_0,\, L_{\pm 1}\] = \mp L_{\pm 1}~~, ~~~~
{}~~~~ \[ L_1,\, L_{-1} \] = 2 L_0~~,
\eqno(3.6) $$
through the identification
$$ T_1 = -2 L_0 ~~, ~~~~ T_2 = L_1 + L_{-1} ~~, ~~~~
T_3 = - L_1 + L_{-1}  ~~.
\eqno(3.7) $$
For our purpose of utilizing EH instanton (3.1), the $~T_I\,$'s in (3.5)
is more advantageous.

        Performing the identifications (3.4) we get the solution for
$~A\du\m I$:
$$\eqalign{& A\du 22 = - \half \sqrtrf \sinhth \cos\psi ~~,
{}~~~~A\du 2 3 = - \half \sqrtrf \sinhth \sin \psi ~~,  \cr
&A\du 2 1 = \half \left( 1 + \fracmm1{r^4} \right) \coshth~~, \cr
&A\du 3 2 = \half \sqrtrf \sin\psi ~~, ~~~~
A\du 3 3 = - \half \sqrtrf \cos\psi ~~, \cr
&A\du 4 1 = \half \left( 1 + \fracmm1{r^4} \right) ~~ , \cr }
\eqno(3.8) $$
and all other components are zero.
The satisfaction of the SD condition (2.6) is easily
confirmed for the field strength
$$F_{\m\n}{}^I = \partial_\m A\du\n I - \partial_\n A\du\m
I + f\du{J K} I A\du\m J A\du\n K ~~.
\eqno(3.9) $$
A remarkable point is that despite of the EH
gravitational instanton background, the SD condition (2.6) {\it does} hold
for
our SDYM instanton.  For explicitness, we give the YM field strength:
$$\eqalign{&F\du{1 2} 1 = - \fracmm 2{r^5} \coshth ~~, ~~~~
F\du{1 4} 1 = - \fracmm2{r^5} ~~, ~~~~ F\du{23} 1 = - \fracmm 1{r^4}
\sinhth ~~, \cr
&F\du{12} 2 = -\fracmm{\sinhth\cos\psi}{r^5\sqrtrf} ~~,
{}~~~~  F\du{23} 2 = - \fracmm{\sqrtrf \coshth \cos\psi}{2r^4} ~~, \cr
&F\du{34} 2 = \fracmm{\sqrtrf \cos\psi}{2r^4} ~~, ~~~~
F\du{24} 2  = \fracmm{\sqrtrf \sinhth\sin\psi}{2r^4} ~~, ~~~~  \cr
&F\du{12} 3 = - \fracmm{\sinhth \sin\psi}{r^5\sqrtrf}~~, ~~~~
F\du{13} 3  = -\fracmm{\cos\psi}{r^5\sqrtrf} ~~, ~~~~  \cr
&F\du{23} 3 = - \fracmm1{2r^4} \sqrtrf \coshth\sin\psi ~~, ~~~~
F\du{24} 3 = - \fracmm1{2r^4} \sqrtrf \sinhth\cos\psi ~~, ~~~~  \cr
&F\du{34} 3 = \fracmm{\sqrtrf \sin\psi}{2r^4} ~~, ~~~~  \cr }
\eqno(3.10) $$
and all other independent components are zero.

        The important ingredient about our prescription above is the
special role played by the $~SL(2)$~ indices $~{\scst m n}$~ in
$~\omega\du\m{m n}$~ or $~R\du{\m\n}{m n}$, as if they were ``internal''
$~SL(2)$~ gauge symmetry.  This is the $~SO(2,2)$~ analog of the usual
Euclidean case $~SO(4) \approx SU(2) \otimes SU(2)$~ [13].

        As for the gaugino equation (2.3), we simply put the gaugino to
zero, similarly to the gravitino case.  This can be done consistently
with supersymmetry, as long as such an ansatz satisfies all the field
equations.  This is also reasonable from the fact that we have already
fixed the freedom of supersymmetry, when the gravitino is put to zero, and
eventually the background has {\it no} manifest supersymmetry.

        Our remaining field equations are for the SDTM.  As in the case
of the gaugino, we can satisfy (2.2) by the trivial solution
$~\chi_\a=0$, like the gravitino and gaugino.  Eq.~(2.5) is rather
easily solved now.  First we rewrite (2.5) in the form
$$\Bo \phi - \fracm 13 F\du{\m\n} I F\ud{\m\n} I = 0 ~~,
\eqno(3.11) $$
where
$$~\phi \equiv \exp\left(\fracm2{\sqrt3}\Phi\right)~~.
\eqno(3.12) $$
Inserting the solution (3.10) into (3.11), we get
$$\phi''(r) + \fracmm{3r^4+1}{r(r^4-1)} \phi'(r) =
\fracmm{32}{r^8(r^4-1)} ~~,
\eqno(3.13) $$
where we have assumed that $~\phi$~ depends {\it only} on $~r$, and
each prime is for the derivative $~d/dr$.  This can be easily
solved by
$$\phi = - \fracmm{2(3r^4+1)}{3r^6} + \fracmm{a - 4}4 \ln\left(
\fracmm{r^2-1}{r^2+1}\right) + 1 ~~,
\eqno(3.14) $$
where we imposed the boundary condition $~\Phi (r\rightarrow \infty) = 0
$, and $~a$~ is an arbitrary constant corresponding to the general
solution of (3.13).

        The $~B_{\m\n}$~ field is
now solved as (3.18) below by the help of (2.5) as follows.
The non-vanishing independent component of
$~G_{\m\n\r}$~ obtained from (2.5) with (3.14) is
$$G_{2 3 4} =  \fracmm {{\sqrt3}(4-ar^8)}{16r^8} \sinhth~~.
\eqno(3.15) $$
Here the CS term played a peculiar role for the integrability
for the potential field $~B_{\m\n}$, as guaranteed by
the Bianchi identity:
$$\partial_{\[ \m} G_{\n\r\s\]} = - \fracm{\sqrt3}2 F\du{\[\m\n} I
F_{\r\s\]\, I} ~~.
\eqno(3.16) $$
valid for our solution (3.10).

To summarize our results, we have the solutions
$$\li{&\Phi = \fracm{\sqrt3}2 \ln\left| 1 - \fracmm{2(3r^4+1)}{3r^6}
+ \fracmm{a-4}4 \ln\left( \fracmm{r^2 - 1}{r^2 + 1} \right)  \right|
{}~~,
&(3.17) \cr
&B_{2 4} = \fracmm{4 + 3a}{16\sqrt3} \coshth + b\,(\varphi,\,\psi) ~~,
&(3.18) \cr  } $$
for the SDTM, and the solution (3.10) or
(3.8) for the SDYM on the EH background (3.1).  The $~b\,(\varphi,\,\psi)$~
is an arbitrary function only of $~\varphi$~ and $~\psi$.

        Due to the non-compactness of our space-time, the various
topological integrals [14] for our solutions do not converge.  This is mainly
caused by the integral $~\int_0^\infty d\vartheta \, \sinhth\rightarrow
\infty$.  The action or hamiltonian (before imposing the
SD conditions [12], or in what we call Parkes-Siegel (PS)
formulation [6,10]) for our exact solutions is also divergent due to a
boundary term with the same $~\vartheta\-$integral.  However,
we expect that an appropriate Wick-rotation can lead our space to the
compact Euclidian space-time, which replaces the above integral by the
{\it finite} one: $~\int_0^{\pi} d \theta \, \sin\theta = 2$.  In
fact, we get the Euler number [14] after such a replacement:
$$\eqalign{\chi (M) = \, & \fracmm1{16\pi^2} \int_{\rm M} d^4 x \,
R^{m n}\wedge R_{m n}
- \fracmm1{16\pi^2} \int_{\partial{\rm M}} \e^{m n r s} \left[ \omega_{m n}
\wedge R_{r s} - \fracm23 \omega_{m n}\wedge\omega\du r t\wedge\omega_{t s}
\right] \cr
= & - \half (-3) - (-\half) = + 2~~.  \cr }
\eqno(3.19) $$
{}From these viewpoints, we regard our exact solutions as equally
important as the Euclidian cases [14].

        Another interesting feature of our solutions is the topological
significance of the $~B_{\m\n}\-$field related to the instanton number
{\it via} the CS term.  In particular, we have
$$\eqalign{\int_{\rm M} & F^I \wedge F_I = - {\sqrt3} \int_{\rm M} d G
= - {\sqrt 3} \int _{\partial{\rm M}} G \cr
& = \fracmm 3 2\int_{\partial{\rm M}}
* d\left(e^{2\Phi/{\sqrt3}}\right)
= \fracmm 3 2\int_{\rm M} d^4 x\, e \Bo\phi  ~~. \cr }
\eqno(3.20) $$
The penultimate equality is due to the SD in the SDTM.

        The next natural question is the compatibility and relationship of
our exact solution with $~N=2$~ string theory.  In our previous paper
[4], we have given a possible Green-Schwarz (GS) $~\s\-$model formulation.
In our
present case we have to rearrange the action, such that the coupling is
consistent with our {\it canonical} fields.  Leaving all the details to
Ref.~[12], we give the Wess-Zumino-Witten (WZW) term in the GS $~\s\-$model
relevant to our discussion here:
$$ I_{\rm WZW} = \int d^2 \s\, \left[- \fracm1{2\pi}\fracm 1{\sqrt3} \,
\e^{i j} \Pi\du i A \Pi \du j B B_{B A}\, \right] ~~,
\eqno(3.21) $$
where $~\Pi\du i A\equiv (\partial_i Z^M)E\du M A$~ with the $~D=(2,2)$~
(inverse)vielbein $~E\du M A$, and the superspace coordinates $~Z^M$.
The indices $~{\scst i,~j,~\cdots~=~1,~2}$~ are for the world-sheet
curved coordinates.  In the absence of fermionic backgrounds
we can replace $~\Pi \du i A$~ by $~(\partial_i X^\m) e\du\m m$~
(Neveu-Ramond-Schwarz $~\s\-$model).  The special factor $~1/{\sqrt3}$~
is for the normalization fixed by our {\it
canonical} fields [12].  To see the effect of $~B_{\m\n}$~ on the WZW-term,
we regard the $~r\-$coordinate as a ``time'' variable for our
``instanton'' solution (3.10).  Considering also the appropriate
normalization factors, the
effect of such instanton at time $~r$~ yields the exponent\footnotew{We have
included the effect of YM gauge anomaly in the CS
term by the GS mechanism [17], when $~3\partial_{\[\m} B_{\n\r\]}$~ is
converted into $~\Hat G_{\m\n\r}$~ in (3.22).  See Ref.~[16] for the details.}
$$\eqalign{ P(r) &= \fracm1{2\pi}\fracm1{\sqrt3} \int d^2 \s \, \e^{i j}
(\partial_i X^\m) (\partial_j X^\n) B_{\m\n} \cr
& = \fracm1{2\pi} \fracm 1{\sqrt3} \int d^2\s \int_0^1 du \, \e^{\hat
i\hat j\hat k} (\partial_{\hat i} \Hat X^\m) (\partial_{\hat j} \Hat X^\n)
(\partial_{\hat k} \Hat X^\r) \Hat G_{\m\n\r} \cr
& = \fracmm{3(4-ar^8)} {16\pi r^8} \int _0^{2\pi} d\varphi \int_0^{2\pi}
d\psi \int_0^\pi d\theta \sin\theta \cr
& = \fracmm{6\pi}{r^8} - \fracmm{3a\pi}2 ~~, \cr}
\eqno(3.22) $$
in the string path-integral [16].  We have used what is called
``Vainberg construction'' [15,16], introducing a third new coordinate
$~0\le u\le 1$.  Accordingly all the quantities
with {\it hats} are associated with the total extended three-dimensional
manifold, and $~\Hat G_{\m\n\r}$~ is a function of $~\Hat X^\m(\s,u)$,
such that $~\Hat X^\m(\s,0) = 0,~\Hat X^\m(\s,1) = X^\m(\s)$~ [15,16].
We have also identified these three-dimensional coordinates
with the $~D=(2,2)$~ ones by $~\s^1 = \varphi,~ \s^2 = \theta,~
u = \psi /(2\pi)$.
Then the second expression of (3.22) contains nothing else than the
Jacobian of this ``coordinate transformation'' from
$~(\s^1,\,\s^2,\,u)$~ to $~(\Hat X^\m)$.
Now the total contribution to the ``phase-shift'' in the string path-integral
by our instanton between the ``time'' interval $~1\le r <\infty$~ is [16]
$$P(\infty) - P(1) = - 6\pi~~.
\eqno(3.23) $$
Since this is a multiple of $~2\pi$, our exact solutions are consistent
with the $~N=2$~ superstring as its background!  Needless to say, we have also
used the above-mentioned replacement (Wick-rotation) for the
$~\vartheta\-$integral.  Interestingly, the integral in (3.22) is
proportional to (3.20).
This result has strong indication of topological
significance and consistency of the $~N=2$~ string theory [4] formulated on
our SDSG + SDSYM + SDTM background.  Notice also that our
computation involves various numerical factors,
showing the powerful usage of our {\it canonical} notation.

        The prescription we utilized above to get the $~SL(2)$~ SDYM
from the SDG solution is universally applied to other cases,
such as the Taub-Nut solution [18].  Some subtlety arises only
when the Lorentz connection $~\omega\du\m{m n}$~
is {\it not} self-dual, even though $~R\du{\m\n}{m n}$~ {\it is}
self-dual.  In such cases, we can always
arrange $~\omega\du\m{m n}$~ by appropriate Lorentz transformation such that
it is manifestly self-dual [14].  For such self-dual $~\omega$'s, the
identification (3.4) is straightforward to get an exact solution
for $~SL(2)~$ SDSYM.

\bigskip\bigskip

\noindent 4.{\it ~~Concluding Remarks.}~~In this Letter we have
presented a set of exact solutions for the coupled system of
$~N=1$~ SDSG + SDSYM + SDTM on the EH gravitational instanton background for
the first time.  In our system, the
dilaton field $~\Phi$~ played a peculiar role as a part of the SDTM,
in particular by the special coupling to the SDSYM {\it via} the
CS term in the third-rank field strength $~G_{\m\n\r}$.
We also stress that our field equations for the SDTM is
required by supersymmetry combined with the SD condition.
Our {\it canonical} set of field equations are important for our
derivations, due to the manifest SD for the Riemann tensor, which was obscure
in other systems such as in Ref.~[4].

        In this Letter we put the fermionic backgrounds to be trivial,
and eventually the background solutions have {\it no} manifest
supersymmetry.  However, we emphasize that
our background exact solutions are {\it consistent} with supersymmetry, in the
sense that they satisfy {\it supercovariant} field equations (2.1) - (2.6).

        Even though our exact solutions are based on the
$~N=1$~ SDSG + SDSYM + SDTM system, they will be
important also in the PS-formulation [6,10] for extended supersymmetries with
some additional multiplier fields, as well as in other formulations
[6,12].  This is because the field equations
for the non-multiplier fields stay exactly the same as our system after
appropriate truncation into $~N=1$~ supersymmetry, being satisfied by the
same exact solutions.  For example, out of the 70 scalars in the
$~N=8$~ SDSG multiplets [10], our SDTM emerges after appropriate
duality-transformations.

        In this Letter we have also checked the consistency of our exact
solutions as a background for $~N=2$~ superstring theory, by inspecting the
contribution to the WZW-term in the string path-integral.
Remarkably our instanton solution contributes only $~-6 \pi i$~ as a
phase-shift after a Wick-rotation, indicating the validity of our solutions as
consistent $~N=2$~ superstring background.  This result reflects
non-trivial {\it topological} aspects of the system, different from
other {\it perturbative} features such as the $~\b\-$functions treated in
Ref.~[4], providing an independent test for our background
together with our GS $~\s\-$model [4] itself.

        To our knowledge, our dilaton solution is the first peculiar
example of its exact solutions, which is directly related to the antisymmetric
tensor $~B_{\m\n}$~ in a non-trivial way, also as the consistent
background for the $~N=2$~ superstring.

        As a by-product of our exact solutions themselves, we gave a general
algorithm for getting an exact solution for SDYM for the gauge group
$~SL(2)$~ out of any known SDG solution, as a modification of the method known
for Euclidean space-times [13].  For example, we can repeat the
same derivation for the Taub-Nut background [18] for the same field
content.

        Another useful application of our results is to the Euclidean
{\it compact} space-time manifold [14], which has more
advantages such as the topological indices or finite actions.
After appropriate Wick-rotation, we can get the Euclidean version of our
exact solutions.

        It is well-known that the SDYM and SDG produce
non-linear integrable field equations [2,7].  It is then
natural to expect
that their {\it supersymmetric} generalizations are also
integrable.\footnotew{Actually in our recent paper [19] we have shown
that $~N=1$~ SDSYM theory will embed $~N=1$~ and
$~N=2$~ supersymmetric KdV equations and $~N=1$~ supersymmetric Toda theory in
two-dimensions, after appropriate dimensional reductions.}~~As a
matter of fact, $~D=(2,2)$~ self-dual gauge theories can be
identified with {\it two-dimensional} non-linear sigma-models on twistor
surfaces with infinite dimensional algebra such as $~\lim_{n\rightarrow
\infty}\, SL(n)$~ [8], and this connection should be
supersymmetrized.  The resulting
gauge symmetries are the area-preserving diffeomorphisms, indicating possible
link to the $~W_{\infty}\-$algebra [9].  Our exact solutions are the first
explicit examples for the SDSG + SDSYM + SDTM, which can connect these
self-dual gauge theories.

\bigskip\bigskip

We are grateful to D.~Depireux, S.J.~Gates, Jr., T.~H{\" u}bsch,
T.~Jacobson, K.~Pirk and W.~Siegel for important discussions.

\vfill\eject

\refs

\items{1} H. Ooguri and C. Vafa, \mpl{5}{90}{1389};
\np{361}{91}{469}; \ibid{367}{91}{83};
\item{  } H.~Nishino and S.J.~Gates, Jr., \mpl{7}{92}{2543}.
\items{2} A.A.~Belavin, A. M. Polyakov, A. Schwartz and Y. Tyupkin,
\pl{59}{75}{85};
\item{  } R.S.~Ward, \pl{61}{77}{81};
\item{  } M.F.~Atiyah and R.S.~Ward, \cmp{55}{77}{117};
\item{  } E.F.~Corrigan, D.B.~Fairlie, R.C.~Yates and P.~Goddard,
\cmp{58}{78}{223};
\item{  } E.~Witten, \prl{38}{77}{121};
\item{  } A.N.~Leznov and M.V.~Saveliev, \cmp{74}{80}{111};
\item{  } L.~Mason and G.~Sparling, \pl{137}{89}{29};
\item{  } I.~Bakas and D.A.~Depireux, \mpl{6}{91}{399}; {\it ibid.} 1561;
2351.

\items{3} S.V.~Ketov, S.J.~Gates and H.~Nishino, Maryland preprint,
UMDEPP 92--163 (February 1992).

\items{4} H. Nishino, S. J. Gates, Jr. and S. V. Ketov,
Maryland preprint, UMDEPP 92--171 (February 1992).

\items{5} S. J. Gates, Jr., H. Nishino and S. V. Ketov, Maryland
preprint, UMDEPP 92--187 (March 1992), to appear in Phys.~Lett.~B.

\items{6} S.J.~Ketov, H.~Nishino and S.J.~Gates, Jr., Maryland preprint,
UMDEPP 92--211 (June 1992), to appear in Nucl.~Phys.~B.

\items{7} M. F. Atiyah, unpublished;
\item{  } R. S. Ward, Phil.~Trans.~Roy.~Lond.~{\bf A315} (1985) 451;
\item{  } N. J. Hitchin, Proc.~Lond.~Math.~Soc.~{\bf 55} (1987) 59.

\items{8} Q.-H.~Park, \pl{238}{90}{287}; \ibid{257B}{91}{105};
\item{  } Cambridge preprint, DAMTP R-91/12 (October, 1991).

\items{9} See, e.g., E.~Sezgin, Texas A \& M preprint, CTP-TAMU-13/92
(February, 1991).

\items{10} W.~Siegel, Stony Brook preprint, ITP-SB-92-31 (July 1992).

\items{11} T.~Eguchi and A.~Hanson, \ap{120}{79}{82};\\
E.~Calabi, Ann.~Sci.~Ec.~Norm.~Sup.~{\bf 12} (1979) 269.

\items{12} H.~Nishino, Maryland preprint, in preparation (October 1992).

\items{13} J.M.~Charap and M.~Duff, \pl{69}{77}{445}.

\items{14} G.W.~Gibbons and S.W.~Hawking, \cmp{66}{79}{291};
\item{  } T.~Eguchi, P.B.~Gilkey and A.J.~Hanson, Phys.~Rep.~{\bf 6C} (1980)
213.

\items{15} S.J.~Gates, Jr.~and H.~Nishino, \pl{173}{86}{46} and 52.

\items{16} R.~Rohm and E.~Witten, \ap{170}{86}{454}.

\items{17} M.~Green and J.H.~Schwarz, \pl{149}{84}{117}.

\items{18} S.W.~Hawking, Phys.~Lett.~{\bf 60A} (1977) 81.

\items{19} H.~Nishino and S.J.~Gates, Jr., Maryland preprint,
UMDEPP 93--51 (September, 1992).

\end{document}